\documentclass[superscriptaddress,twocolumn,prl,showpacs,floatfix]{revtex4}
\usepackage[dvips]{graphicx}
\usepackage{amsmath,amsfonts,amssymb,bm,ulem}
\usepackage{graphicx}

\begin{document}
\title{
Magnetoresistivity and Acoustoelectronic Effects in a Tilted
Magnetic Field in $p$-Si/SiGe/Si Structures with an Anisotropic $g$
Factor}

\author{I.L.~Drichko}
\affiliation{A.F. Ioffe Physico-Technical Institute of Russian
Academy of Sciences, 194021 St. Petersburg, Russia}
\author{I.Yu.~Smirnov}
\affiliation{A.F. Ioffe Physico-Technical Institute of Russian
Academy of Sciences, 194021 St. Petersburg, Russia}
\author{A.V.~Suslov}
\affiliation{National High Magnetic Field Laboratory, Tallahassee,
FL 32310, USA}
\author{O.A.~Mironov}
\affiliation{Warwick SEMINANO R$\&$D Centre, University of Warwick
Science Park, Coventry CV4 7EZ, UK}
\author{D.R. Leadley}
\affiliation{Department of Physics, University of Warwick, Coventry,
CV4 7AL, UK}
\date{\today}
\begin{abstract}
Magnetoresistivity $\rho _{xx}$ and $\rho _{xy}$ and the
acoustoelectronic effects are measured in $p$-Si/SiGe/Si with an
impurity concentration $p$ = 1.99 $\times $ 10$^{11}$ cm$^{-2}$ in
the temperature range 0.3--2.0 K and an tilted magnetic field up to
18 T. The dependence of the effective $g $factor on the angle of
magnetic field tilt $\theta $ to the normal to the plane of a two
dimensional $p$ Si/SiGe/Si channel is determined. A first order
ferromagnet-paramagnet phase transition is observed in the magnetic
fields corresponding to a filling factor $\nu $ = 2 at $\theta
\approx $ 59$^\textrm{o}$-60$^\textrm{o}$.

\end{abstract}

\pacs{73.23.-b, 73.43.Qt} \maketitle

\section{Introduction} \label{intr}

The rather interesting kinetic properties of $p$-Si/SiGe/Si
structures have been attracting attention of researchers for the
last two decades [1-6]. The quantum well in $p$-Si/Si$_{1-
x}$Ge$_{x}$/Si is located in the strained Si$_{1- x}$Ge$_{x}$ layer;
therefore, the triply degenerate (without regard for spin) valence
band of SiGe splits into three bands due to spin--orbit coupling and
mechanical stresses. The charge carriers are represented by heavy
holes, whose band is formed by atomic states with quantum numbers
$L$ = 1, $S$ = 1/2, and $J$ = 3/2. This should result in a strong
anisotropy of the effective $g$ factor: $g$* $\approx $ 4.5 if a
magnetic field is normal to the plane of the quantum well and $g$*
$\approx $ 0 if a magnetic field is parallel to the plane of the
quantum well [7].

In such structures in a magnetic field normal to the plane of the
quantum well, the values of parameter $m$* and $g$* factor are such
that the relationship $g$*$\mu _{B}B \approx $ $\hbar$$\omega _{c}$
holds true; here, $\mu _{B}$ is the Bohr magneton, $\omega _{c}$ =
\textit{eB}/$m$*$c$ is the cyclotron frequency, $e$ is the electron
charge, $B$ is the magnetic field, $m$* is the effective hole mass,
and $c$ is the velocity of light. As a result, resistivity $\rho
_{xx}$ oscillations only with odd filling factors $\nu $ = 1, 3, 5,
{\ldots} are experimentally observed in magnetic field $B$. $\rho
_{xx}(B)$ oscillations at $\nu $ = 2 are observed only in some
samples. The specific features of conductivity in the magnetic field
corresponding to $\nu $ = 2 are comprehensively studied in this
work.

The formation of pinned Wigner glass was revealed in $p$-Si/SiGe/Si
with a low impurity concentration ($p $= 8.2 $\times $ 10$^{10}$
cm$^{-2})$ in the quantum limit in a perpendicular magnetic field
[8], and a giant positive magnetoresistivity was observed in this
structure in a parallel magnetic field [9]. This magnetoresistivity
was caused by the interaction of the orbital motion of carriers in a
wide quantum well with a parallel magnetic field.

An unusual phenomenon, namely, the so called re-entrant
metal-insulator transition, was also found in this structure in a
magnetic field at a filling factor $\nu $ = 3/2 [1-6]. In [2], this
anomaly was related to the presence of smooth large scale potential
fluctuations with an amplitude comparable with the Fermi energy. In
[3-5], however, this phenomenon was explained by the crossing of
Landau levels with opposite spin directions (0$\uparrow $,
1$\downarrow )$ when the magnetic field strength changes.

In this work, we study magnetoresistivity, Hall effect, and
acoustoelectronic effects in an tilted magnetic field to determine
the dependence of the $g$* factor on the angle of magnetic field
tilt to the normal to the plane of a two dimensional channel. With
this dependence, we can analyze the possible cause of the anomalies
in conductivity that appear at a filling factor $\nu $ = 2 in a $p
$Si/SiGe/Si sample with $p$ = 2 $\times $ 10$^{11}$ cm$^{-2}$. The
asymmetric quantum well in this sample was 30 nm wide, and the
sample structure was described in [9].

\section{EXPERIMENTAL RESULTS} \label{experiment}

Magnetoresistivity $\rho _{xx}$ and $\rho _{xy}$ and the
acoustoelectronic effects were measured in an tilted magnetic field
up to 18 T in the temperature range 0.3-2.0 K in a linear current
regime ($I$ = 10 nA). A sample with an impurity concentration $p$ =
2 $\times $ 10$^{11}$ cm$^{-2}$ and a mobility of 7 $\times $
10$^{3}$ cm$^{2}$/V s was studied. The integer quantum Hall effect
was observed in the sample in a magnetic field normal to the plane
with two dimensional conductivity (Fig. 1).
\begin{figure}[ht]
\centerline{
\includegraphics[width=\columnwidth]{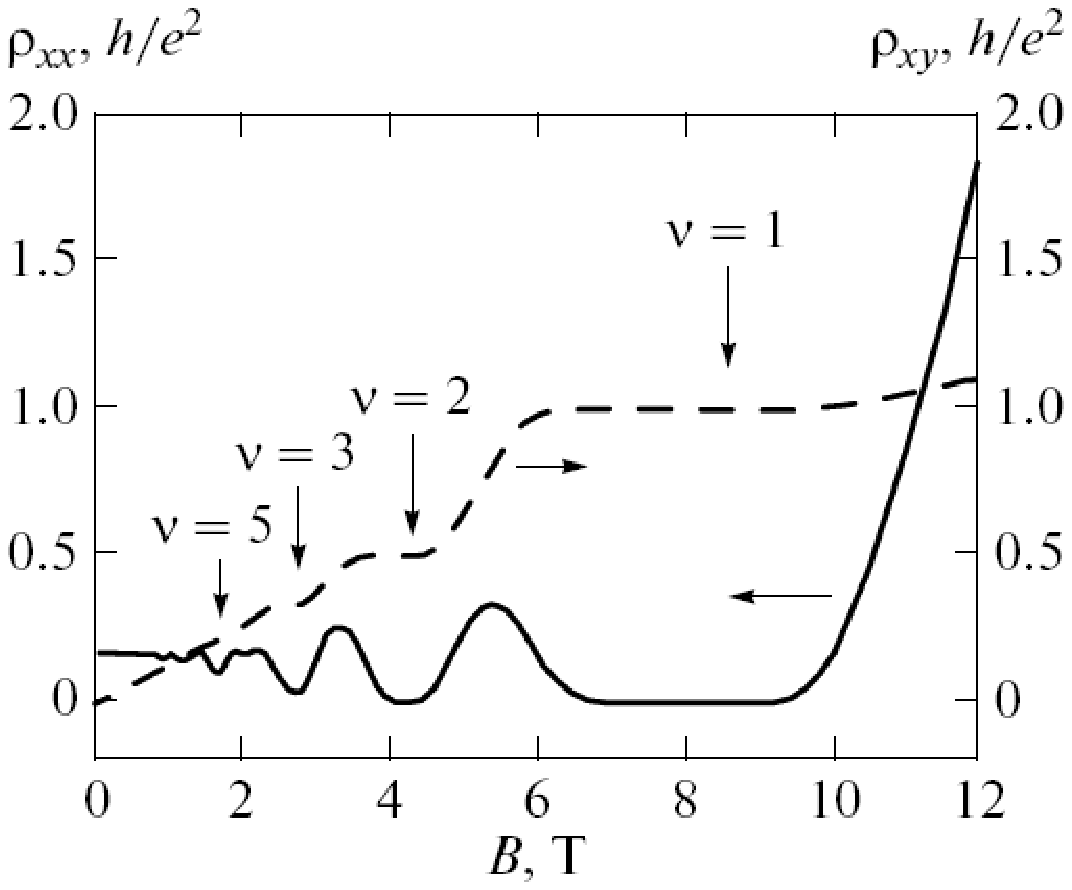}
} \caption{$\rho _{xx}$ and $\rho _{xy}$ (in $h$/$e^{2})$ vs. magnetic field at
$T$ = 0.35 K and $p$ = 1.99 $\times $ 10$^{11}$ cm$^{-2}$.
\label{Fig1}}
\end{figure}

Figure 2a shows $\rho _{xx}$ versus total magnetic field $B $at
various angles of inclination $\theta $ to the normal to the two
dimensional layer surface at $T$ = 0.3 K. At $\theta $ = 0\r{ } in
field $B \approx $ 5.2 T, which corresponds to $\nu $ = 3/2 the
maximum of $\rho _{xx}$ is observed. It is seen that, as $\theta $
increases, the oscillation maxima and minima shift toward high
magnetic fields and the value of $\rho _{xx}$ in the maxima at $\nu
$ = 3/2 increases. Figure 2b shows the $\rho _{xx}(B_{\bot })$
dependence near $\nu $ = 3/2 for various $\theta $. Indeed, the
anomaly described in the Introduction does occur near $B$ = 5.2 T
($\nu $ = 3/2): $\rho _{xx}$ increases more than five fold when
angle $\theta $ increases from 0\r{ } to 70\r{ }. Figure 3 shows the
dependences of $\rho _{xy}$ on the total magnetic field and its
perpendicular component at $T$ = 0.3 K and various angles of
inclination $\theta $.

The conductivity was calculated at various angles of inclination $\theta $,

$\sigma_{xx}$=$\rho_{xx}$/($\rho_{xx}^2$+$\rho_{xy}^2$).

The $\sigma _{xx}(B_{\bot })$ dependence is shown in Fig. 4. It is
inter esting that anomalies do not occur in the $\sigma
_{xx}(B_{\bot })$ dependences at $\nu $ = 3/2 ($\sigma _{xx}$
changes by 30{\%} when tilt angle $\theta $ changes from 0\r{ } to
75\r{ }). Moreover, the positions of the $\sigma _{xx}$ minima in a
magnetic field are inde pendent of the tilt angle at $\nu $ = 1, 3,
and 5, whereas these minima begin to shift toward low magnetic
fields at $\nu $ = 2 and $\theta  \quad >$ 50\r{ }. The explanation
of this fact requires separate consideration.

Apart from the dc measurements, we also studied the
acoustoelectronic effects in an tilted magnetic field at $T $= 0.3
K. The acoustic methods based on sur face acoustic waves make it
possible to measure sam ples of a square or rectangular shape
without application of electric contacts. To this end, a ``hybrid''
method is used: a surface acoustic wave propagates along the surface
of the piezoelectric insulator LiNbO$_{3}$, and the structure to be
studied is slightly pressed to the lithium niobate surface by a
spring. A deformation wave propagating along the lithium nio bate
surface is accompanied by an electric field wave having the same
frequency. The electric field penetrates into the channel with two
dimensional conductivity, inducing high frequency currents and,
correspondingly, wave energy absorption. In this experimental
configuration, deformation is not transmitted into the sample.
\begin{figure}[ht]
\centerline{
\includegraphics[width=\columnwidth]{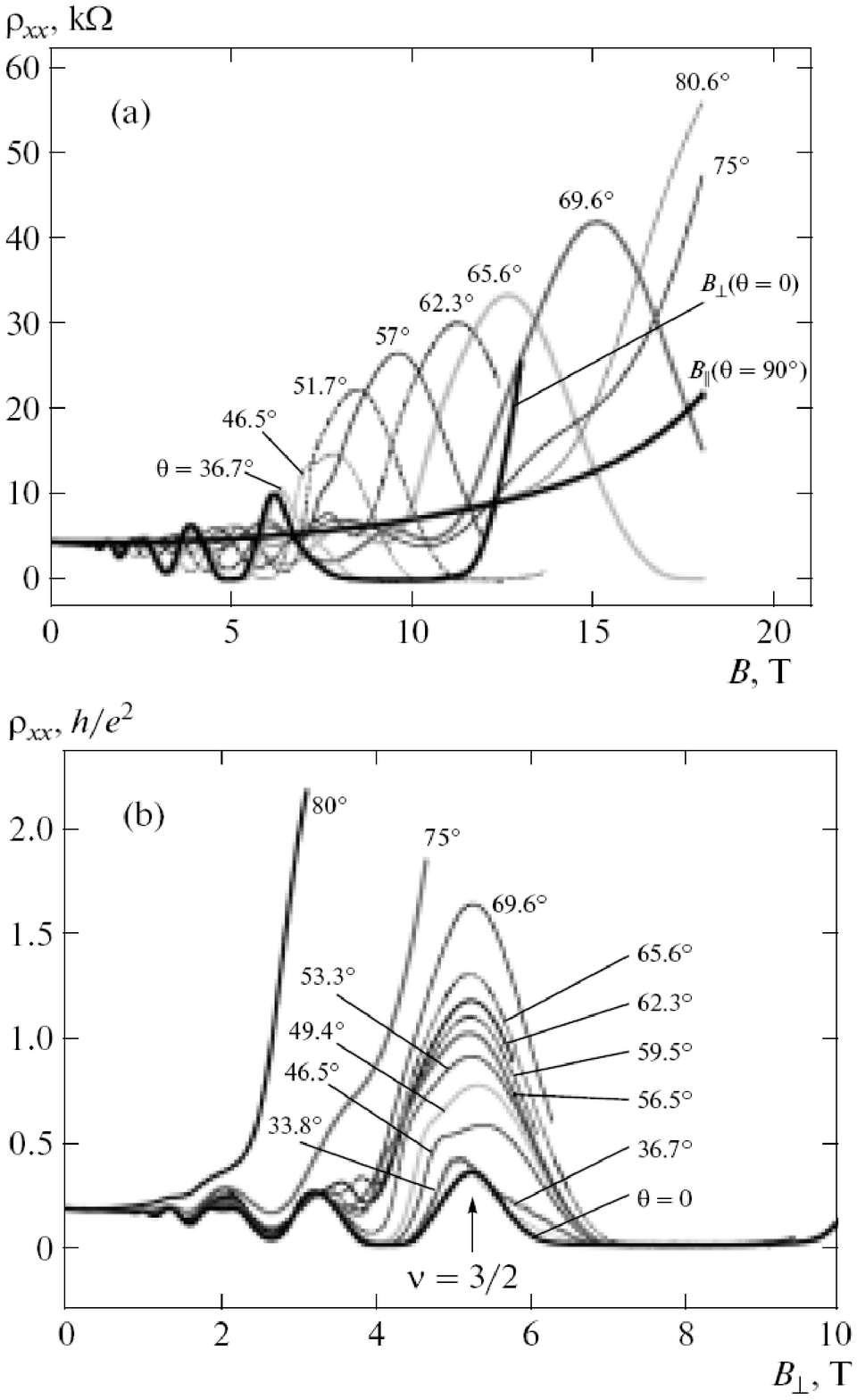}
} \caption{(a) $\rho _{xx}$ vs. total magnetic field at various angles of
inclination $\theta $ to the normal to the sample surface at $p$ = 1.99 $\times
$ 10$^{11}$ cm$^{-2 }$and $T$ = 0.3 K ($B_{\bot }$ and $B_{\vert \vert }$ are the
field components normal and parallel to the structure plane, respectively).
(b) $\rho _{xx}(B_{\bot })$ dependences at various angles $\theta $ and
$T$ = 0.3 K.
\label{Fig2}}
\end{figure}

\begin{figure}[ht]
\centerline{
\includegraphics[width=\columnwidth]{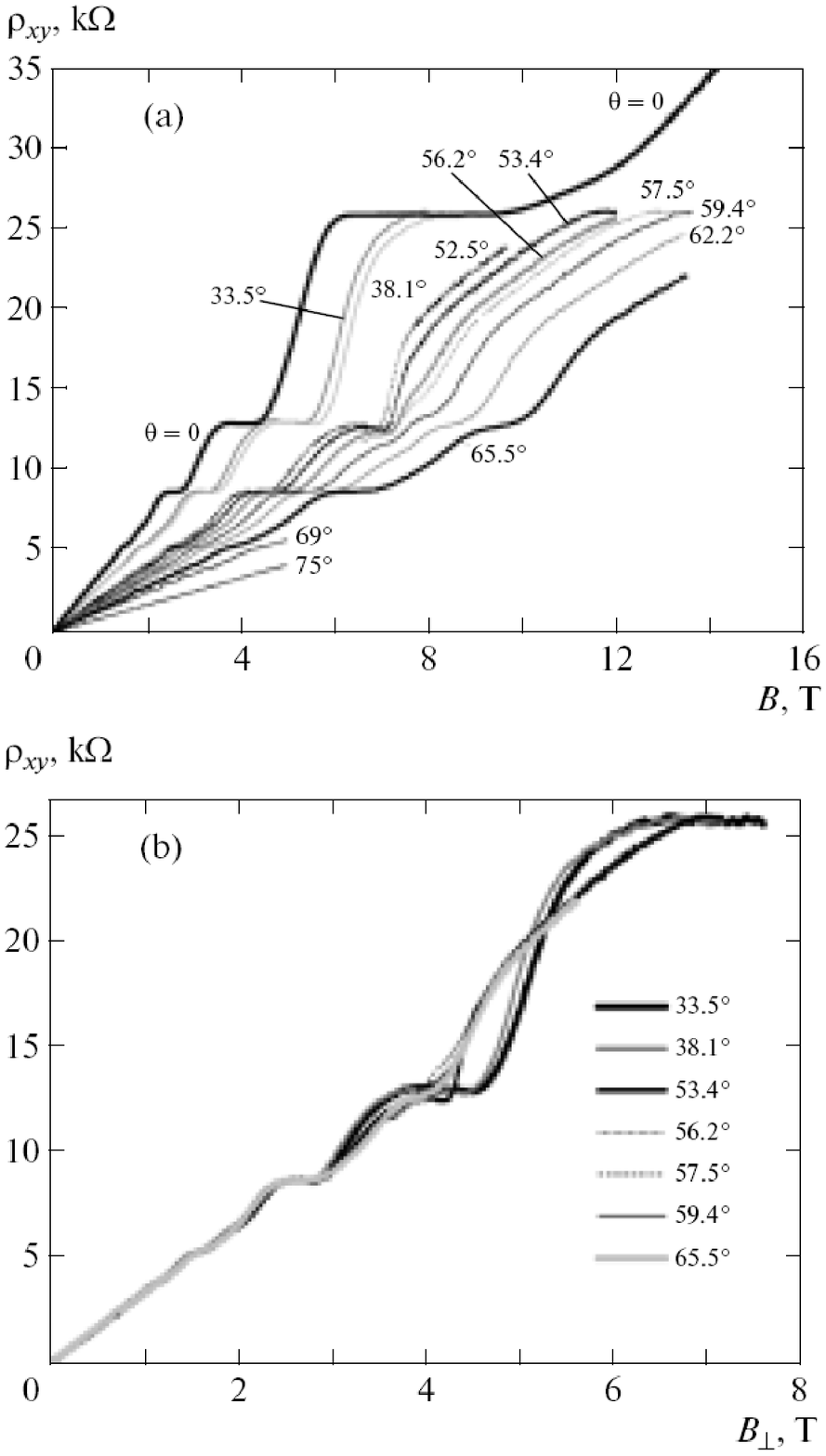}
} \caption{$\rho _{xy}$ as a function of (a) a total magnetic field and
(b) its normal component at $T$ = 0.3 K and various angles of
magnetic field tilt to the normal to the sample surface.
\label{Fig3}}
\end{figure}

\begin{figure}[ht]
\centerline{
\includegraphics[width=\columnwidth]{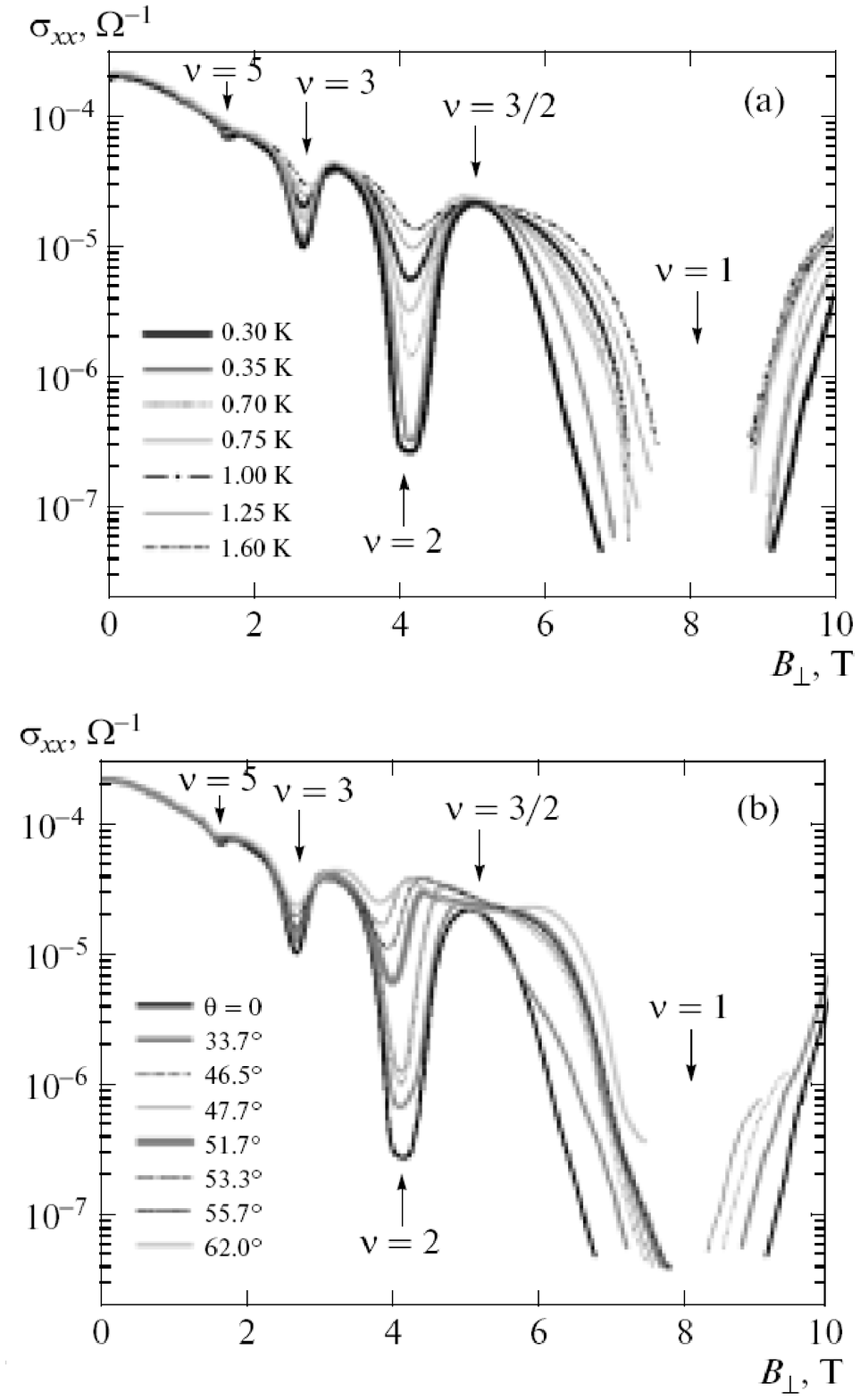}
} \caption{$\sigma _{xx}(B)$ dependences (a) at various temperatures and
$\theta $ = 0 and (b) $\sigma _{xx}(B_{\bot })$ at various angles $\theta
$ and $T$ = 0.3 K.
\label{Fig4}}
\end{figure}

We measured the following acoustoelectronic effects: absorption
coefficient $\Gamma $ and the change in the surface acoustic wave
velocity at a frequency of 30 MHz in an tilted magnetic field up to
18 T for various angles of inclination at $T$ = 0.3 K. The
acoustoelectronic effects are governed by the high frequency
conductivity of a two dimensional system [10]. If the system
exhibits the quantum Hall effect and dc Shubnikov-de Haas
oscillations in a magnetic field, similar oscillations are also
observed in the acoustoelectronic effects.

Figure 5 shows absorption coefficient $\Gamma $ and the change in
the surface acoustic wave velocity $\Delta V$/$V$ versus the total
magnetic field at various angles of magnetic field tilt. It is seen
that the positions of oscillations shift toward high magnetic fields
as the tilt angle increases and that the oscillation corresponding
to $\nu $ = 2 disappears at $\theta $ = 53\r{ }. This fact means
that, at $\theta $ $\approx$ 53\r{ }, the conductivity in the two
dimensional channel increases strongly.

\section{DISCUSSION OF THE EXPERIMENTAL RESULTS}

If an effective $g$ factor is isotropic, its value in an tilted
magnetic field is usually determined by the formula
\begin{equation}
  \label{1}
  g^* = \sqrt{g_{\bot }^{2 }\cos^{2 }\theta +g_{\vert \vert
}^{2}\sin^{2}\theta} ,
\end{equation}

\begin{figure}[ht]
\centerline{
\includegraphics[width=\columnwidth]{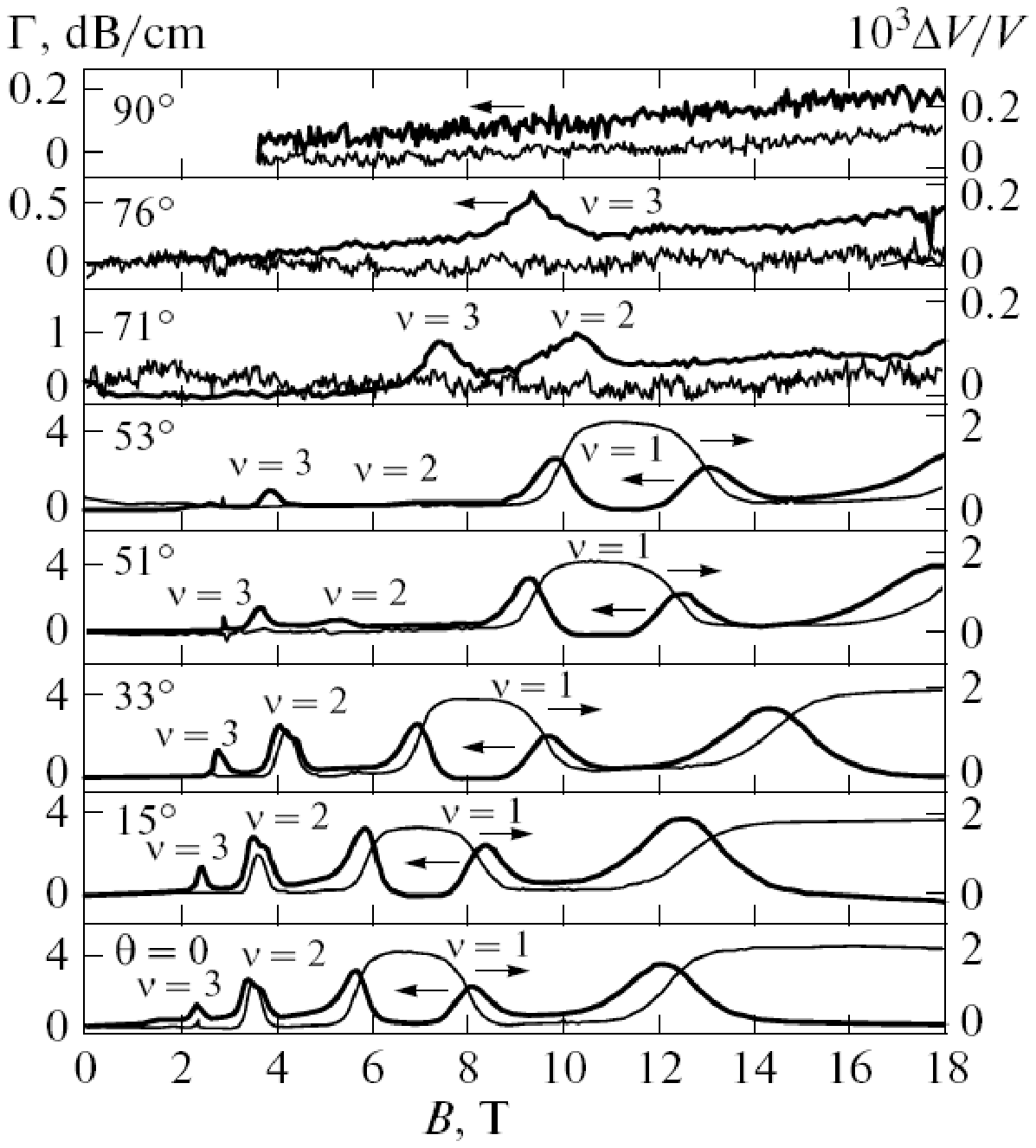}
} \caption{Absorption coefficient $\Gamma $ and the change in the surface
acoustic wave velocity $\Delta V$/$V$ vs. the total magnetic field at various
angles of magnetic field tilt $\theta$ to the normal to the structure
surface, $T$ = 0.3 K, and $f$ = 30 MHz.
\label{Fig5}}
\end{figure}
where $g_{\bot }$is the $g$ factor in the magnetic field normal to
the plane of the channel with two dimensional conductivity and
$g_{\vert \vert }$ is the $g $factor in the magnetic field parallel
to this plane. For a strong anisotropy, where $g_{\vert \vert }$=0
(as should be in the structure under study), we have $g$*= $g_{\bot
}$cos$\theta $. In this case, when a magnetic field is inclined, the
positions of the oscillation minima shift toward high magnetic
fields but the oscillation amplitude is independent of the angle of
magnetic field tilt. Such a situation took place in [11]. In our
samples, the oscillation amplitude depends on angle $\theta$ at all
filling factors; therefore, axial symmetry which is corresponded to
Eq. (1) is likely to be violated in our sample. We use the
dependences of conductivity $\sigma _{xx}$ on temperature and angle
of magnetic field tilt $\theta$ at $\nu $= 3 in order to determine
the angular dependence of the $g$* factor. As a function of the
magnetic field at $\nu $= 3, $\sigma _{xx }$is minimal when the
Fermi level is located between two spin split Landau levels,
1$\downarrow $and 1$\uparrow $. As is seen from Fig. 6, $\sigma
_{xx}$ at $\nu $ = 3 grows with both temperature ($\theta $= 0) and
angle of magnetic field tilt $\theta (T $= 0.3 K). In this case, the
temperature and angular dependences of the conductivity minimum can
be expressed as

$\sigma _{xx }\propto $exp [-$g$* ($\theta) \mu_{B}B$/2\textit{kT}],

where $B$ is the total magnetic field and $k$ is the Boltzmann
constant. Therefore, if we plot the $\sigma _{xx}(T)$ and
$\sigma_{xx}(\theta )$ dependences and attribute certain temperature
$T $' to every angle $\theta$ at the same $\sigma _{xx}$, we can
obtain the expression
\begin{equation}
  \label{freqpin}
  \frac{g^* (\Theta = 0)}{0.3}  = \frac{g^*(\Theta)}{T'}\, ,\nonumber
\end{equation}
or
\begin{equation}
  \label{freqpin2}
  \frac{g^*(\Theta)}{g^*(\Theta=0)}=\frac{T'}{0.3}\, ,\nonumber
\end{equation}
Thus, we can determine the dependence of the reduced $g$* factor on tilt angle
$\theta $. A similar procedure can also be performed for $\nu $ = 5 but with
a much lower accuracy.

Another method for determining $g$*($\theta )$ is related to the
construction of the dependences of $\sigma _{xx}$ on total magnetic
field $B$ for the oscillation minima at $\nu $= 3, 5, 7, and 9;
various angles $\theta $; and$ T$ = 0.3 K. The conductivities were
determined from the curves shown in Fig. 2. Since $\sigma _{xx
}\propto $exp(-$g$*($\theta )\mu _{B}B$/2\textit{kT}) as noted
above, the ln$\sigma _{xx}(B)$ dependences at various $\theta $are
described by linear functions. In this case, we have $\Delta
$(ln$\sigma _{xx})$/$\Delta B \quad \sim  \quad g$*($\theta )$;
therefore, we can determine the ratio $g$*($\theta )$/$g$*($\theta $
= 0). As is seen from Fig. 7, the slope increases with decreasing
angle $\theta $. It turns out that the values of $g$*($\theta
)$/$g$*($\theta $= 0) determined by different methods at the same
value of $\theta $are similar; they are shown as points in Fig. 8 at
$g$*($\theta $= 0) = 6 ($g$*($\theta $= 0) is determined below).
Figure 8 (curve) also shows the $g$*($\theta )$ = 6cos$\theta $
dependence. As follows from Fig. 8, the $g$*($\theta )$ dependence
in the sample under study differs from the curve calculated by Eq.
(1).

We now consider the $\sigma _{xx}(B_{\bot })$ dependences of
conductivity near $\nu $ = 2 for various temperatures and angles
$\theta $(Figs. 9, 10).

As noted above, only odd minima are usually observed in Shubnikov-de
Haas oscillations in experiment in Si/SiGe/Si samples in a magnetic
field nor mal to the plane of a two dimensional channel. The only
exception is the even minimum at $\nu $= 2, which is only observed
in some samples, in particular, in the sample under study. It is
generally accepted that the absence of oscillations with an even
filling factor $\nu $is caused by the relation $\hbar$$\omega _{c
}\approx g$*$\mu _{B}B$. When this equality holds true in magnetic
fields corresponding to even $\nu $, the energies of the levels with
different spin directions are the same in different Landau bands;
therefore, oscillations with even $\nu $ are not observed. The value
of $g$* calculated from this equality at $m$* = 0.21$m_{0}$ is 4.7.
However, as was shown in our experiment, we both observed a clear
minimum of $\sigma _{xx}$ at $\nu $= 2 (Fig. 9a) and were able to
measure its temperature dependence. The $\sigma _{xx}(T)$ dependence
at $\theta $ = 0 corresponds to an activation energy $E_{a }\approx
$3.2 K (0.28 meV), which means that levels 0$\uparrow $and
1$\downarrow $at $\theta $= 0 are close but noncoincident.
\begin{figure}[ht]
\centerline{
\includegraphics[width=\columnwidth]{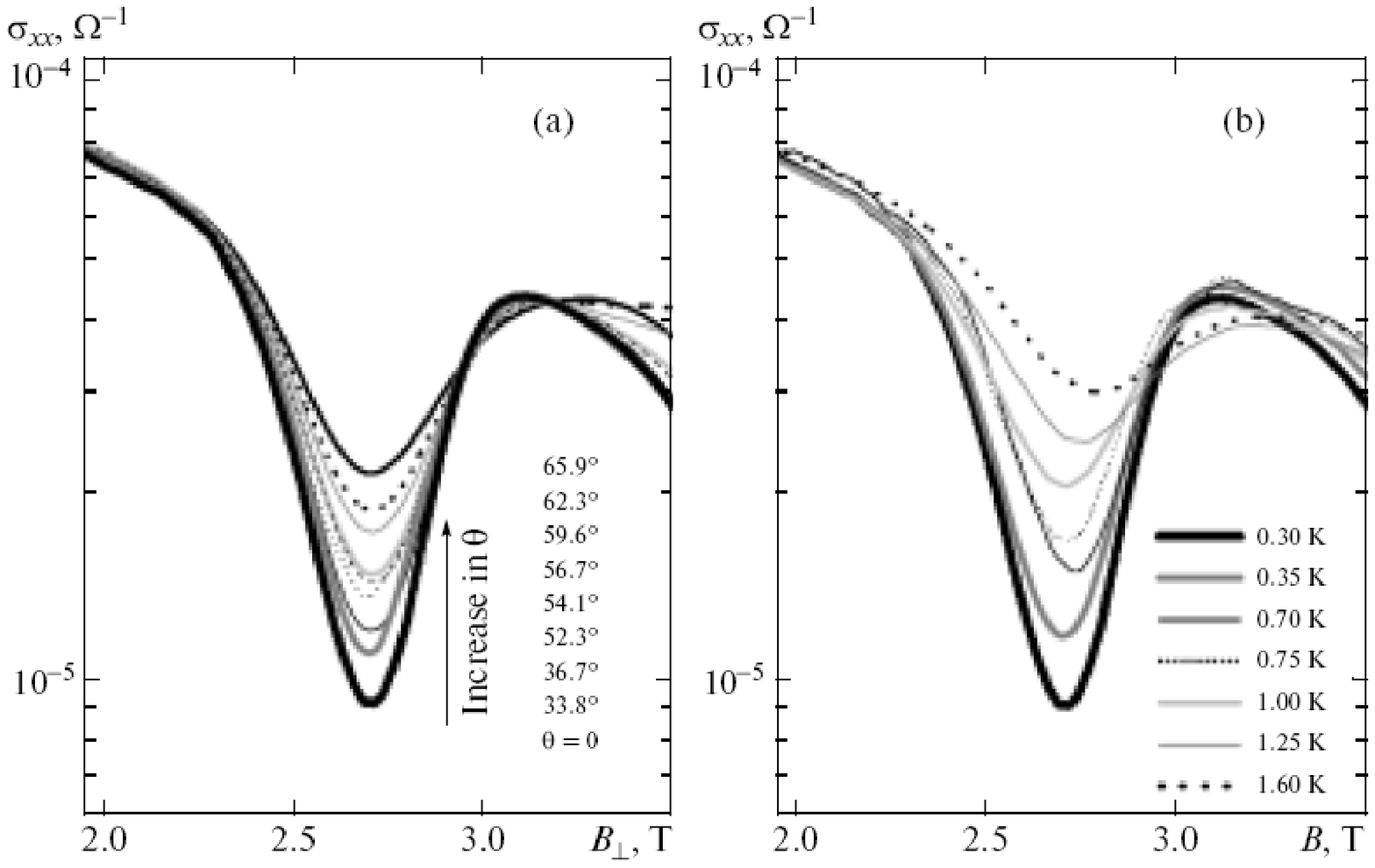}
} \caption{$\sigma _{xx}(B_{\bot })$ dependences (a) at various angles of
inclination $\theta $, $\nu $ = 3, and $T$ = 0.3 K and (b) $\sigma
_{xx}(B_{\bot })$ at various temperatures,$\nu \quad =\quad 3,$ and $\theta $ = 0.
\label{Fig6}}
\end{figure}

\begin{figure}[ht]
\centerline{
\includegraphics[width=\columnwidth]{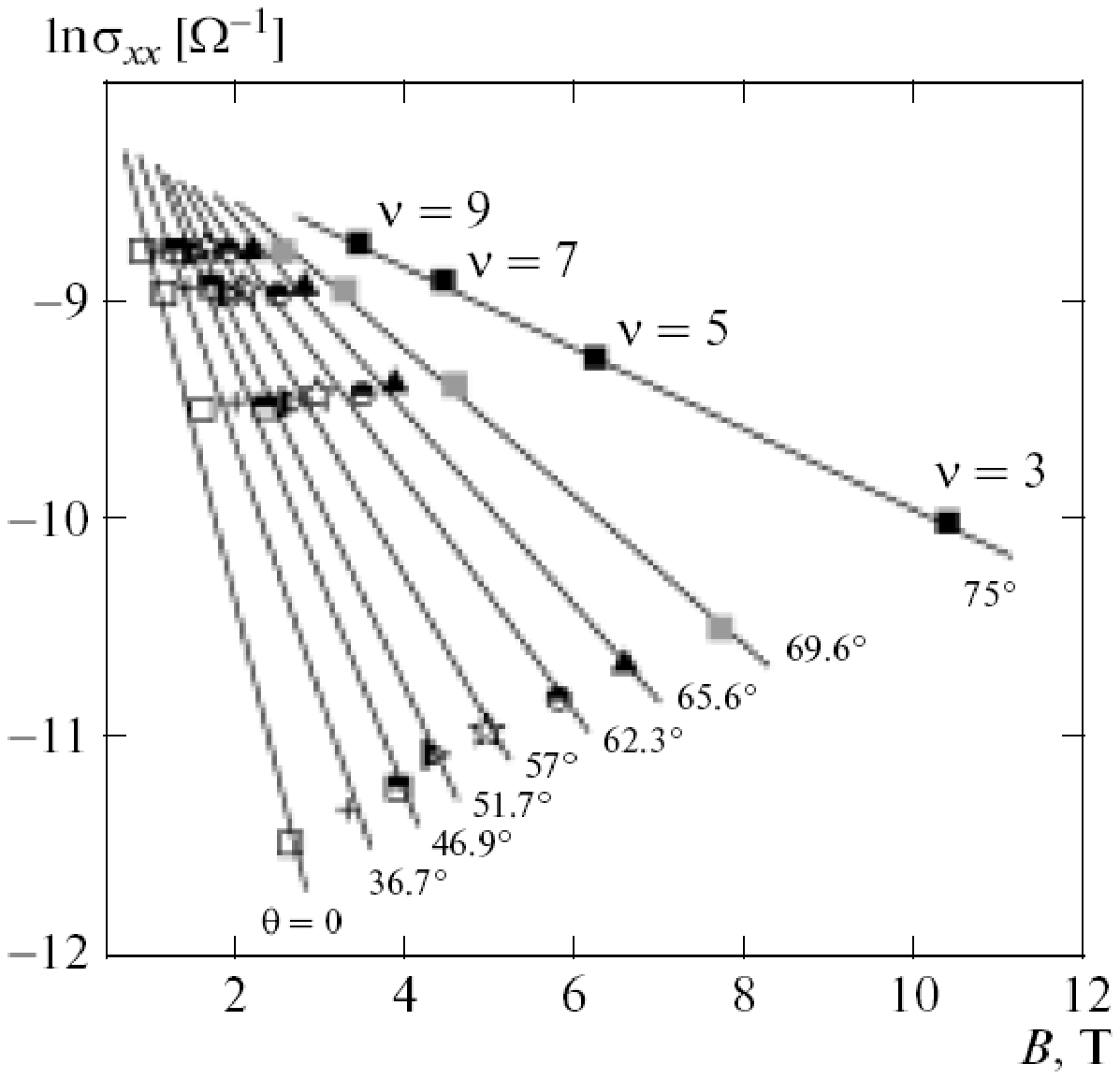}
} \caption{ln$\sigma _{xx}$ vs. magnetic field at $\nu $ = 3, 5, 7, and 9 and
various angles of inclination $\theta $.
\label{Fig7}}
\end{figure}

It is also seen from Fig. 9 that conductivity $\sigma _{xx}$ in the
minimum at $\nu $ = 2 and a varied magnetic field increases with
both the temperature and angle. When the temperature changes, the
position of this oscillation minimum does not change along the
magnetic field axis (Fig. 9a); however, as tilt angle $\theta$
increases, the conductivity minimum shifts toward low magnetic
fields (Fig. 9b). When the tilt angle reaches about 59.5\r{ } (Fig.
10), the following two oscillations are present in the curve: the
former oscillation with its minimum shifting to the left with
increasing angle (at $\theta \approx $ 59.5\r{ }, it is observed at
$B_{\bot }$ = 3.6 T) and a new oscillation appearing at $B_{\bot
}\approx $ 4 T. As $\theta $ increases further, the amplitudes of
the new oscillations increase and their minima shift toward low
magnetic fields, and the former oscillations disappear. Both types
of oscillations coexist in the angular range 59.5\r{ } $\le \theta
\le $ 61\r{ }.

\begin{figure}[ht]
\centerline{
\includegraphics[width=\columnwidth]{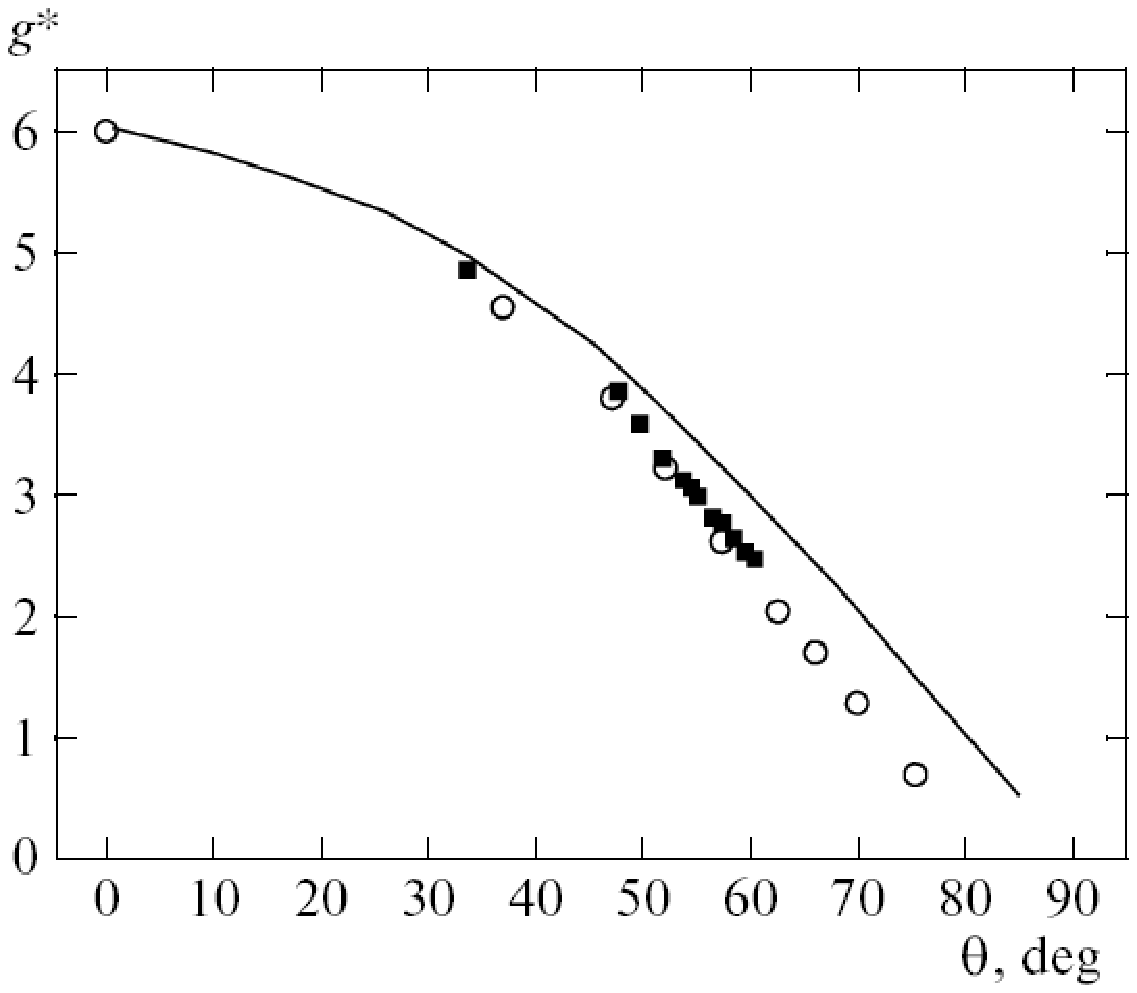}
} \caption{Dependence of the $g$* factor on the angle of mag netic field tilt to
the normal to the sample surface: (curve) $g$*($\theta )$ = 6 cos$\theta $,
(circles) experimental data at $\nu $ = 3, and (squares) experimental data
at $\nu $ = 2.
\label{Fig8}}
\end{figure}

Figure 11 shows the $\sigma _{xx}(B_{\bot })$ dependences in the
field range 2.5-5.2 T for various angles $\theta $ shifted at a step
of 5 $\times $ 10$^{-6 }\Omega ^{-1}$ with respect to the curve with
$\theta $ = 55.7\r{ } (for clarity).

To analyze these anomalies, we use the idea employed in [3-5] to
explain the specific features of electrical resistivity in
Si/SiGe/Si at $\nu $ = 3/2. They were related to the change of the
relative position of spinsplit (0$\uparrow $, 1$\downarrow )$ Landau
levels induced by the dependence of the $g$* factor of the magnetic
field. This idea was developed earlier in theoretical works [12, 13]
and was supported experimentally in GaAs/AlGaAs for $\nu $ = 4 [14]
and 2/3 [15]. It was assumed that the crossing of the Landau levels
is accompanied by a first order paramagnet-ferromagnet phase
transition [14, 15].
\begin{figure}[ht]
\centerline{
\includegraphics[width=\columnwidth]{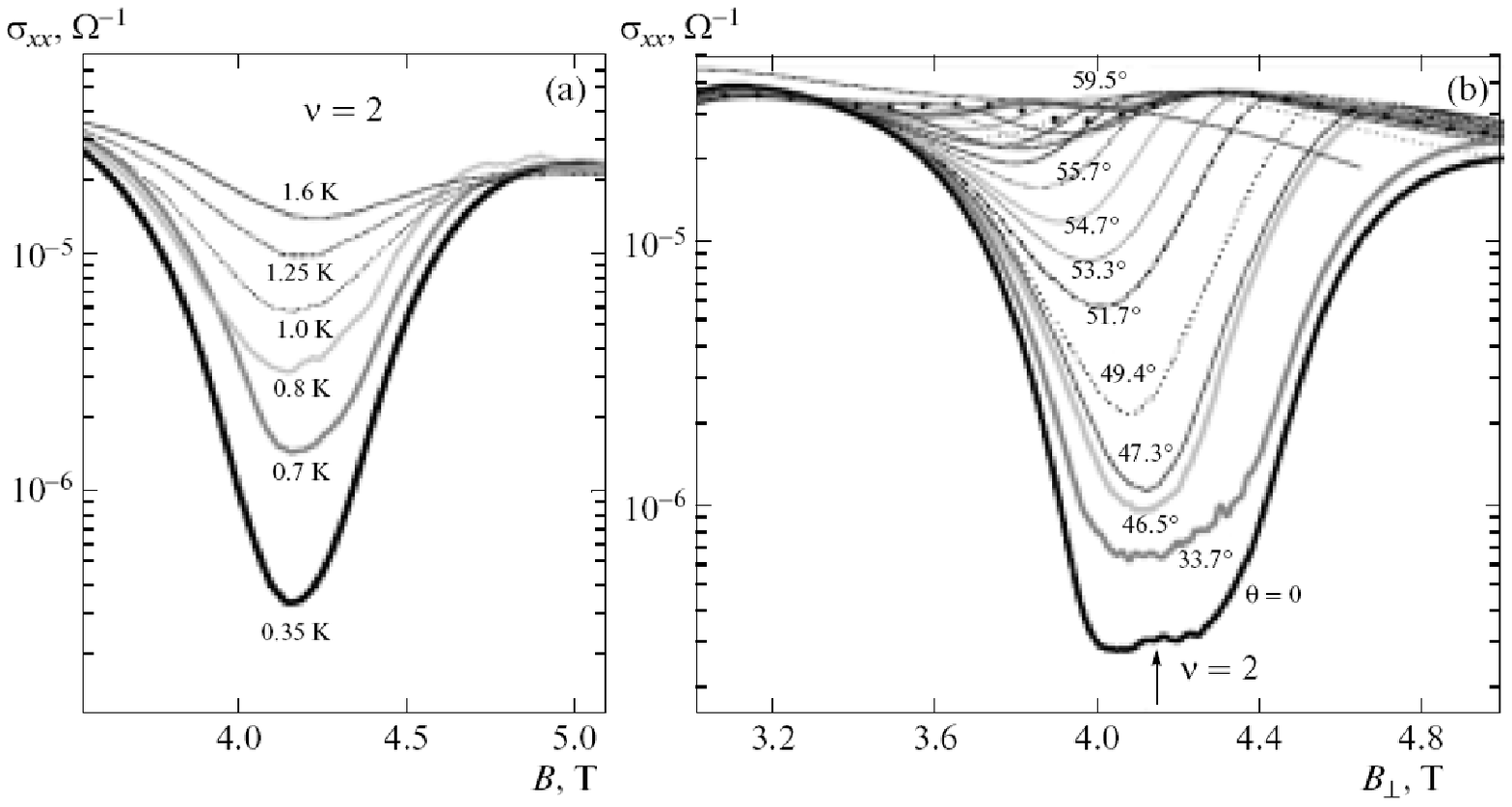}
} \caption{$\sigma _{xx}(B)$ dependences near $\nu $ = 2 at (a) various
temperatures and $\theta $ = 0 and (b) $\sigma _{xx}(B_{\bot })$ near
$\nu $ = 2 at angles $\theta $ = 0\r{ }--59.5\r{ }and $T$ = 0.3 K.
\label{Fig9}}
\end{figure}

We now study the relative position of Landau levels 0$\uparrow $ and
1$\downarrow $ at $\nu $ = 2 ($B_{\bot }$ = 4.15 T) as a function of
the angle of magnetic field tilt. To this end, we need to know the
value of $g$*($\theta $ = 0). We choose $g$*($\theta $ = 0) and
$m$*/$m_{0}$ so that their product is $g$*($\theta $ = 0)$m$*/$m_{0
}\approx $ 2 (which follows from the equality $\hbar$$\omega _{c
}\approx \quad g$*$\mu _{B}B)$ and obtain $g$*($\theta $ = 0) = 6
and $m$* = 0.32$m_{0}$ (where $m_{0 }$is the free electron mass). We
now use the$ g$*($\theta )$/$g$*($\theta $ = 0) dependence, which is
obtained experimentally and shown in Fig. 8, and the experimental
value of the activation energy (0.28 meV) and construct the
dependences of levels 0$\uparrow $ and 1$\downarrow $ on angle
$\theta $ using the formulas
\begin{eqnarray}
  \label{2}
 && E(0\uparrow)=\hbar\omega_c/2 + g^*\mu_B B/2, \,   \nonumber \\
 && E(1\downarrow)=3\hbar\omega_c/2 - g^*\mu_B B/2, (2) \nonumber
\end{eqnarray}

\begin{figure}[ht]
\centerline{
\includegraphics[width=\columnwidth]{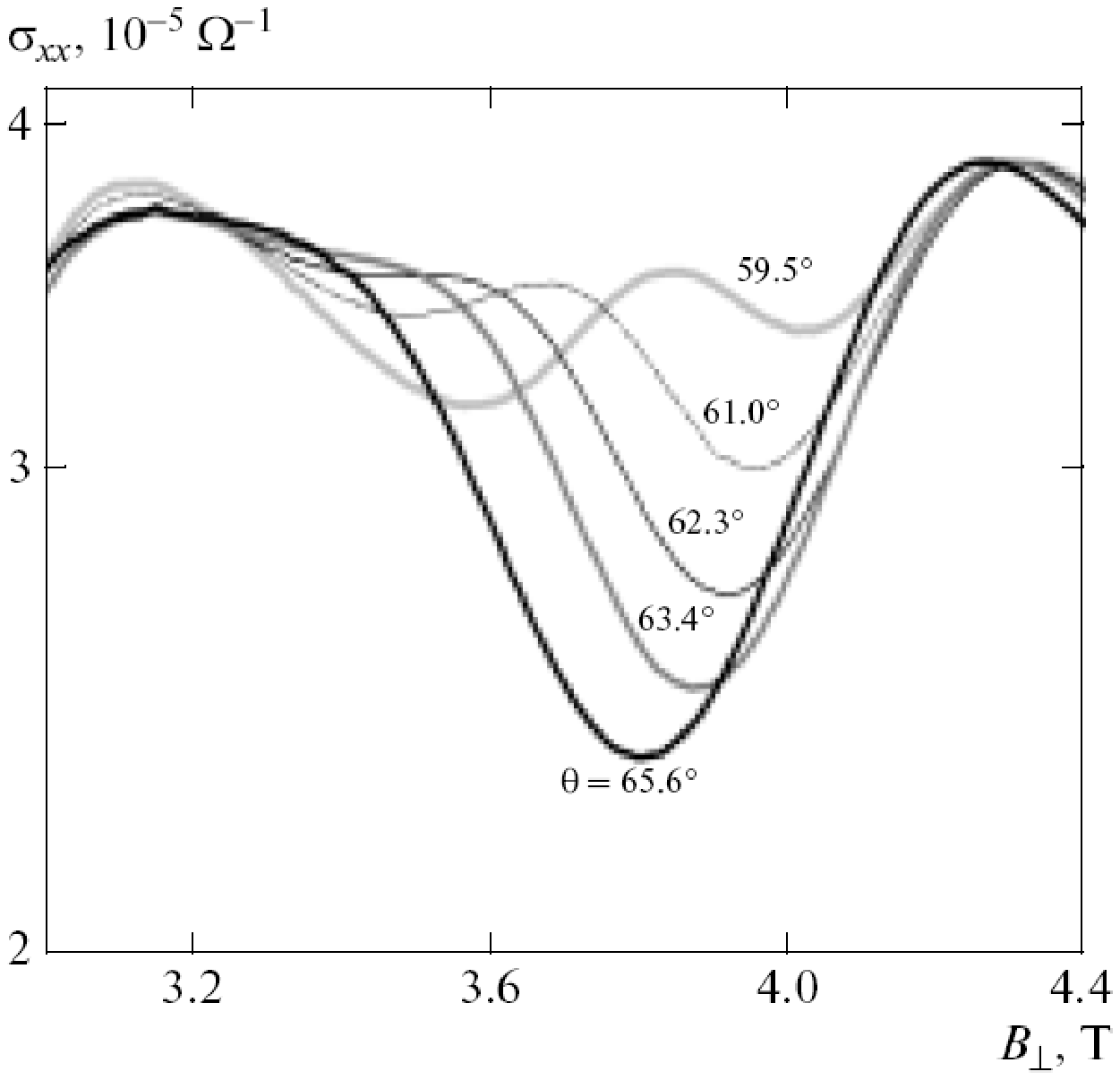}
} \caption{$\sigma _{xx}(B_{\bot })$ dependences at various angles $\theta
$.
\label{Fig10}}
\end{figure}

When choosing the values of $g$* and $m$*, we take into account that
these levels should cross at $\theta $ = 60\r{ }, since the
anomalies are observed experimentally at this angle (see Fig. 11).
The calculation results are shown in Fig. 12. It should be noted
that this calculation is a rough illustration of the possibility of
the crossing of these levels. Here, we do not take into account the
interaction between the levels, the disorder induced broadening of
the levels, and the carrier distribution nonuniformity in a quantum
well [16].

It should be noted that the $g$*($\theta )$/$g$*($\theta $ = 0)
dependence was determined from the dependences of the spin gap
energy on temperature and angle $\theta $ at $\nu $ = 3. However, at
a large spin splitting and $\nu $ = 3, the order in the relative
position of the Landau levels can also be broken. In this case, the
energy determined in the angular range 0\r{ }$\le
 \quad \theta \le $ 60\r{ } differs from the spin gap. As a check, we also
determined $g$*($\theta )$ from the $\sigma _{xx}(\theta )$ dependence at
$\nu $ = 2 in the angular range 0\r{ }$\le  \quad \theta \le $ 60\r{ } (see
the inset to Fig. 13). Since the dependence
\begin{eqnarray}
  \label{3}
\sigma_{xx} \propto \exp \frac{\hbar\omega_c-g^*\mu_B B}{2kT},  \nonumber
\end{eqnarray}
holds true in this angular range, the $\sigma _{xx}(\theta $=
0)/$\sigma _{xx}(\theta )$ ratio can be represented by the
equation
\begin{eqnarray}
  \label{4}
&&\frac{\sigma_{xx}(\theta=0))}{\sigma_{xx}(\theta))} =  \nonumber \\
&& = \exp \frac{-g^*(\theta=0)\mu_B B+g^*(\theta)\mu_B B}{2kT},
\end{eqnarray}
which can be used to determine $g$*($\theta )$. The values of $g$*
determined from this equation are shown as squares in Fig. 8. It is
seen that the values of $g$*($\theta )$ determined by different
methods are close, which means that the crossing of the Landau
levels weakly distorts the spin gap at $\nu $= 3.

\begin{figure}[ht]
\centerline{
\includegraphics[width=\columnwidth]{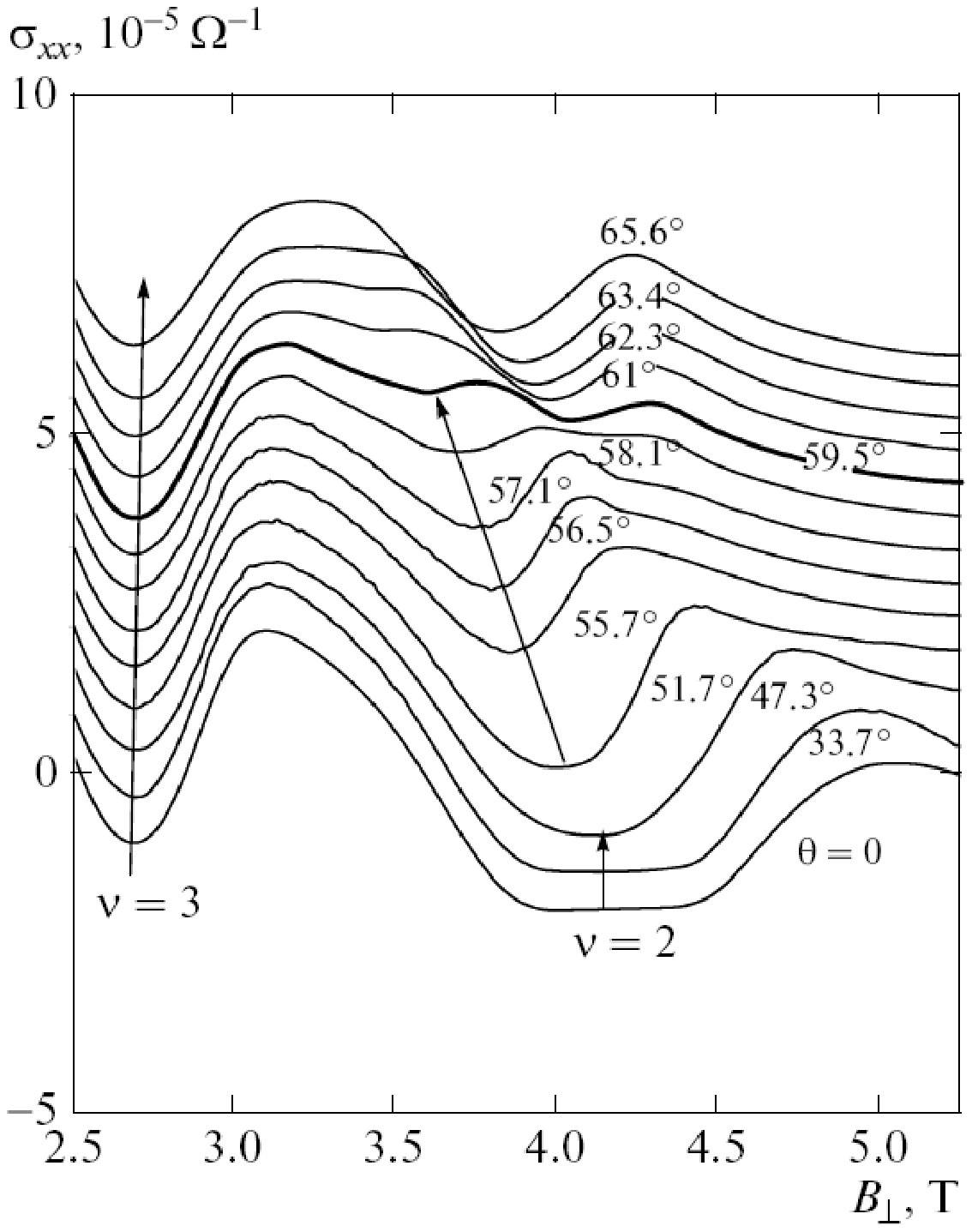}
} \caption{$\sigma_{xx} (B)$ dependences at various magnetic field tilt angles.
For clarity, the curves are spaced 5 $\times $ 10$^{-6}\Omega ^{-1 }$ apart
with respect to the curve at $\theta $ = 57.1\r{ }
(arrows indicate the positions of the minima).
\label{Fig11}}
\end{figure}

For the Landau levels to be intersected, we have to assume (as noted
above) that the energy of level 0$\uparrow $ is higher than that of
level 1$\downarrow $in a magnetic field normal to the two
dimensional layer, i.e., at $\theta $= 0. This means that a
ferromagnetic order should be in the system, since two filled Landau
levels with the same spin direction lie below the Fermi level. At
$\theta \approx $60\r{ }, the levels intersect each other and, at
$\theta >$ 60\r{ }, change their relative position, breaking the
ferromagnetic order. In the angular range from 59.5\r{ }to 61\r{ },
both states coexist.

\begin{figure}[ht]
\centerline{
\includegraphics[width=\columnwidth]{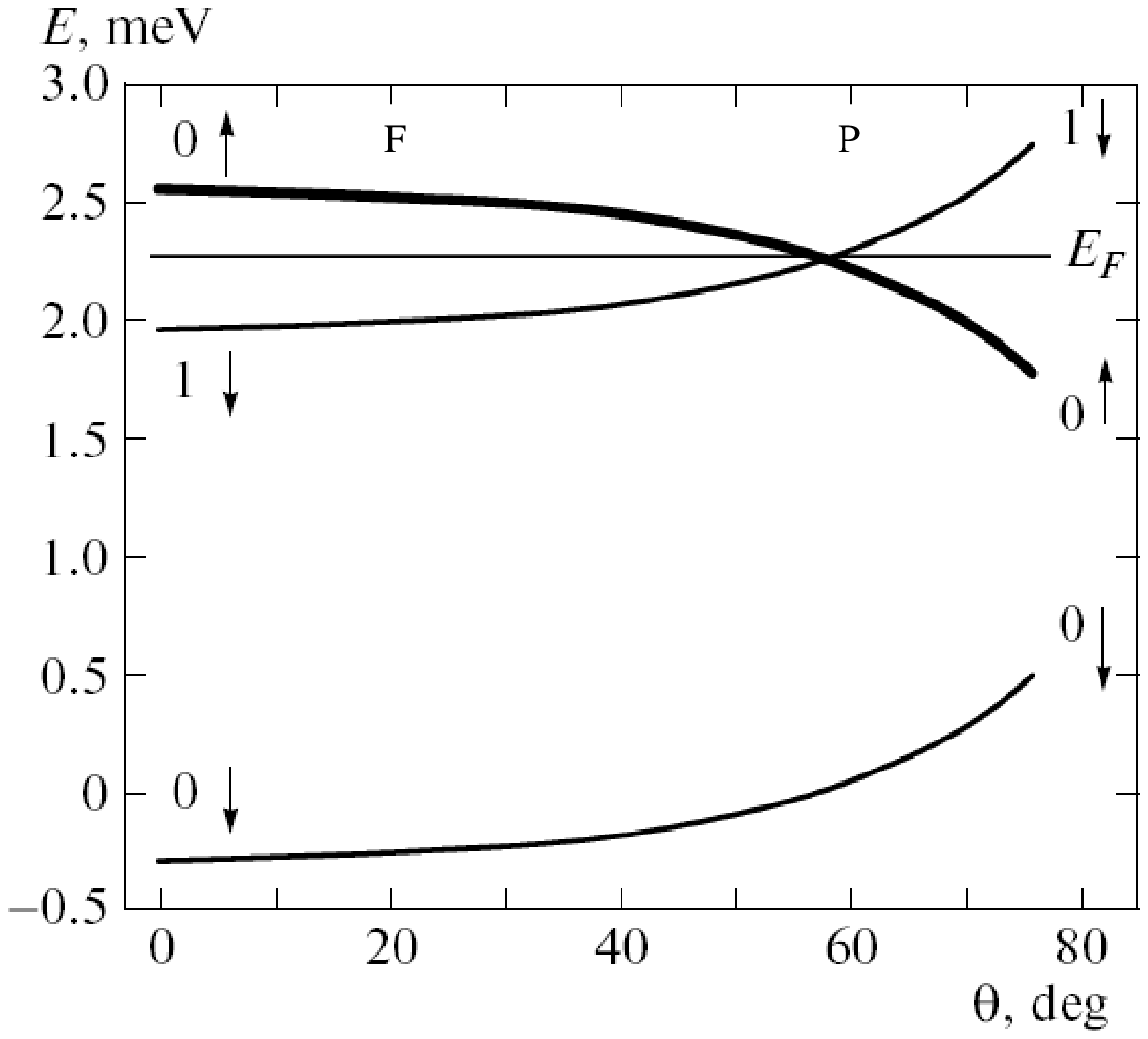}
} \caption{Position of Landau levels with respect to Fermi
level $E_F$ as a function of angle $\Theta$. ($F$ and $P$ stand for ferro- and paramagnet, respectively).
\label{Fig12}}
\end{figure}

Landau levels 0$\uparrow $ and 1$\downarrow $ intersect due to a
strong dependence (decrease) of the $g$* factor on the angle of
magnetic field tilt. At the point of crossing, we have $g$* $\approx
$ 2.37. In this case, the value of filling factor $\nu $calculated
from the position of the oscillation minima in the magnetic field
axis changes jumpwise. The $\nu (\theta )$ dependence is shown in
Fig. 13. Both phases are seen to coexist in the angular range
59.5\r{ } $\le \theta \le $61\r{ }. However, the conductivity at the
oscillation minima does not jump and reaches a maximum at the
transition point. This dependence can easily be explained, since the
activation energy at the transition point is minimal and increases
on either side of this transition. This fact is also supported by
the measurement of the acoustoelectronic effects, whose magnitude is
governed by the conductivity. The dependences of the
acoustoelectronic effects on the magnetic field and the tilt angle
support the conclusion about the absence of the anomaly in
conductivity at $\nu $ = 3/2.

\begin{figure}[ht]
\centerline{
\includegraphics[width=\columnwidth]{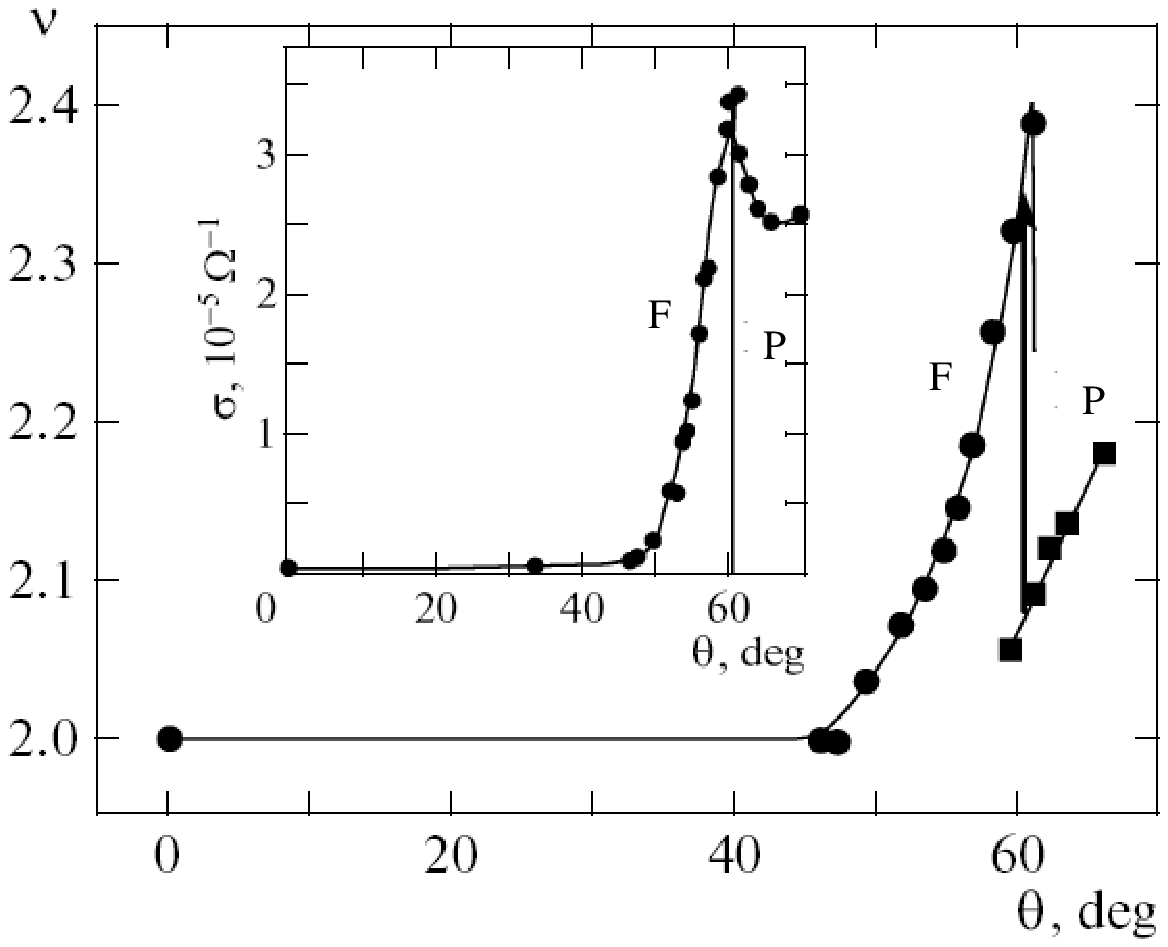}
} \caption{Filling factor $\nu $ vs. angle $\theta $. The inset shows
the dependence of conductivity $\sigma _{xx}$ at oscillation minima
on angle $\theta $.
\label{Fig13}}
\end{figure}

In conclusion, we would like to note the following.

\begin{enumerate}
\item When analyzing a number of works dealing with the resistivity anomaly in a magnetic field at $\nu $ = 3/2, we can conclude that the properties of $p$Si/SiGe/Si samples are ambiguous: in some samples, the resistiv ity anomaly exists and depends on the angle of mag netic field tilt $\theta $; in other samples, this anomaly is absent; and in samples of a third group, this anomaly is independent of the tilt angle [3]. We think that the resistivity anomaly should be studied at $\nu $ = 2, since point $\nu $ = 3/2 has no physical meaning at a large spin splitting.
\item The ambiguity of the results obtained on differ ent $p$Si/SiGe/Si samples is thought to be related to different dependences of the$ g$* factor on the angle of magnetic field tilt and, apparently, different absolute values of this factor, which is controlled by the quality of the sample and the hole concentration.
\end{enumerate}

\begin{center}
CONCLUSIONS
\end{center}

We studied the magnetoresistivity and the Hall effect of a
$p$-Si/SiGe/Si structure with a quantum well and an anisotropic $g$*
factor in an tilted magnetic field.

At $\nu $ = 2 and 3, we determined the dependence of the $g $factor
on the angle of magnetic field tilt to the normal to the plane of a
two dimensional channel and found the absolute value of the $g$*
factor for $\theta $ = 0.

A high value of the $g$* factor at $\theta $ = 0 was shown to result
in the fact that, in the magnetic field corresponding to $\nu $ = 2,
the energy of Landau level 0$\uparrow $ is higher than the energy of
level 1$\downarrow $ and fully spin-polarized levels 0$\downarrow $
and 1$\downarrow $ lie below the Fermi level. This leads to a
ferromagnetic order. When the angle of mag netic field tilt changes,
the $g$* factor decreases, the normal order of the Landau levels is
restored at $\theta \approx $ 59.5\r{ }, and the ferromagnetic order
is broken. A first order ferromagnet-paramagnet phase transition
occurs, which is indicated by a sharp jump in filling factor $\nu $
and the coexistence of two phases at the transition point.

\begin{center}
ACKNOWLEDGMENTS
\end{center}

I.L.D and I.Yu.S are grateful to Yu.M. Gal'perin, S.A. Tarasenko,
L.E. Golub, V.T. Dolgopolov, E.V. Devyatov, R.V. Parfen'ev, and V.A.
Sanina for useful discussions. We thank T. Murphy and E. Palm for
their assistance in the experiments.

This work was supported by the Russian Foundation for Basic Research
(project no. 080200852); Presidium of the Russian Academy of
Sciences; pro gram Spintronics, Department of Physical Sciences,
Russian Academy of Sciences; and the state program for support of
leading scientific schools (project no. NSh2184.2008.2). A portion
of the work was performed at the National High Magnetic Field
Laboratory (Tallahassee, FL, USA) which was supported by the
National Scientific Foundation (NSF), coopera tive agreement no.
DMR0654118), by the State of Florida.

\begin{center}
REFERENCES
\end{center}

\begin{enumerate}
\item S. I. Dorozhkin, Pis'ma Zh. Eksp. Teor. Fiz. 60 (8), 578 (1994) [JETP Lett. 60 (8), 595 (1994)].
\item S. I. Dorozhkin, C. J. Emeleus, T. E. Whall, G. Landwehr, and O. A. Mironov, Pis'ma Zh. Eksp. Teor. Fiz. 62 (6), 511 (1995) [JETP Lett. 62 (6), 534 (1995)].
\item P. T. Coleridge, A. S. Sachrajda, P. Zawadzki, R. L. Williams, and H. Lafontaine, Solid State Commun. 102, 755 (1997).
\item M. R. Sakr, Maryam Rahimi, S. V. Kravchenko, P. T. Coleridge, R. L. Williams, and J. Lapointe, Phys. Rev. B: Condens. Matter 64, 161 308 (2001).
\item P. T. Coleridge, Solid State Commun. 127, 777 (2003).
\item R. B. Dunford, E. E. Mitchel, R. G. Clark, V. A. Stadnik, F. F. Fang, R. Newbury, R. H. McKenzie,R. P. Starrett, P. J. Wang, and B. S. Meyerson, J. Phys.: Condens. Matter 9, 1565 (1997).
\item E. Glaser, J. M. Trombetta, T. A. Kennedy, S. M. Prokes, O. J. Glembocki, K. L. Wang and C. H. Chern, Phys. Rev. Lett. 65, 1247 (1990).
\item I. L. Drichko, A. M. Dyakonov, I. Yu. Smirnov, A. V. Suslov, Y. M. Galperin, V. Vinokur, M. Myronov, O. A. Mironov, and D. R. Leadley, Phys. Rev. B: Con dens. Matter 77, 085 327 (2008).
\item I. L. Drichko, I. Yu. Smirnov, A. V. Suslov, O. A. Mironov, and D. R. Leadley, Phys. Rev. B: Condens. Matter 79, 205 310 (2009).
\item I. L. Drichko, A. M. Dyakonov, I. Yu. Smirnov, Yu. M. Galperin, and A. I. Toropov, Phys. Rev. B: Condens. Matter 62, 7470 (2000).
\item R. W. Martin, R. J. Nicholas, G. J. Rees, S. K. Haywood, N. J. Mason, and P. J. Walker, Phys. Rev. B: Condens. Matter 42, 9237 (1990).
\item G. F. Giuliani and J. J. Quinn, Phys. Rev. B: Condens. Matter 31, 6228 (1985).
\item S. Yarlagadda, Phys. Rev. B: Condens. Matter 44, 13101 (1991).
\item A. J. Daneshvar, C. J. B. Ford, M. J. Simmons,A. V. Khaetskii, A. R. Hamilton, M. Pepper, and D. A. Ritchie, Phys. Rev. Lett. 79, 4449 (1997).
\item V. T. Dolgopolov, E. V. Deviatov, V. S. Khrapai, D. Reuter, A. D. Wieck, A. Wixforth, K. L. Campman, and A. C. Gossard, Phys. Status Solidi B 243, 3648 (2006).
\item L. Yu. Shchurova, Ann. Phys. (Berlin) 18, 928 (2009).
\end{enumerate}

\end{document}